\documentclass[twocolumn,bibnotes,superscriptaddress,showpacs,amsmath,amssymb,aps,prl]{revtex4-2}

\usepackage{graphicx}
\usepackage{amsmath}
\usepackage{amssymb}
\usepackage{color}
\usepackage{comment}
\usepackage{bm}

\usepackage[colorlinks=true,linkcolor=blue,citecolor=blue]{hyperref}

\begin{document}

\title{Lifting of gap nodes by disorder in ultranodal superconductor candidate FeSe$_{1-x}$S$_x$}

\author{T.~Nagashima}
\author{K.~Ishihara}\email{k.ishihara@edu.k.u-tokyo.ac.jp}
\author{K.~Imamura}
\author{M.~Kobayashi}
\author{M.~Roppongi}
\affiliation{Department of Advanced Materials Science, University of Tokyo, Kashiwa, Chiba 277-8561, Japan}
\author{K.~Matsuura}
\affiliation{Department of Advanced Materials Science, University of Tokyo, Kashiwa, Chiba 277-8561, Japan}
\affiliation{Department of Applied Physics, University of Tokyo, Bunkyo-ku, Tokyo 113-8656, Japan}
\author{Y.~Mizukami}
\affiliation{Department of Advanced Materials Science, University of Tokyo, Kashiwa, Chiba 277-8561, Japan}
\affiliation{Department of Physics, Tohoku University, Aoba-ku, Sendai 980-8578, Japan}
\author{R.~Grasset}
\author{M.~Konczykowski}
\affiliation{Laboratoire des Solides Irradi{\' e}s, CEA/DRF/IRAMIS, Ecole Polytechnique, CNRS, Institut Polytechnique de Paris, F-91128 Palaiseau, France}
\author{K.~Hashimoto}
\author{T.~Shibauchi}\email{shibauchi@k.u-tokyo.ac.jp}
\affiliation{Department of Advanced Materials Science, University of Tokyo, Kashiwa, Chiba 277-8561, Japan}
\begin{abstract}
The observation of time-reversal symmetry breaking and large residual density of states in tetragonal FeSe$_{1-x}$S$_x$ suggests a novel type of ultranodal superconducting state with Bogoliubov Fermi surfaces (BFSs). Although such BFSs in centrosymmetric superconductors are expected to be topologically protected, the impurity effect of this exotic superconducting state remains elusive experimentally. Here, we investigate the impact of controlled defects introduced by electron irradiation on the superconducting state of tetragonal FeSe$_{1-x}$S$_x$ ($0.18\le x\le 0.25$). The temperature dependence of magnetic penetration depth is initially consistent with a model with BFSs in the pristine sample. After irradiation, we observe a nonmonotonic evolution of low-energy excitations with impurity concentrations. This nonmonotonic change indicates a transition from nodal to nodeless, culminating in gapless with Andreev bound states, reminiscent of the nodal $s_\pm$ case. This points to the accidental nature of the possible BFSs in tetragonal FeSe$_{1-x}$S$_x$, which are susceptible to disruption by the disorder.
\end{abstract}
\maketitle

%introduction

\begin{figure*}[t]
    \includegraphics[width=0.85\linewidth]{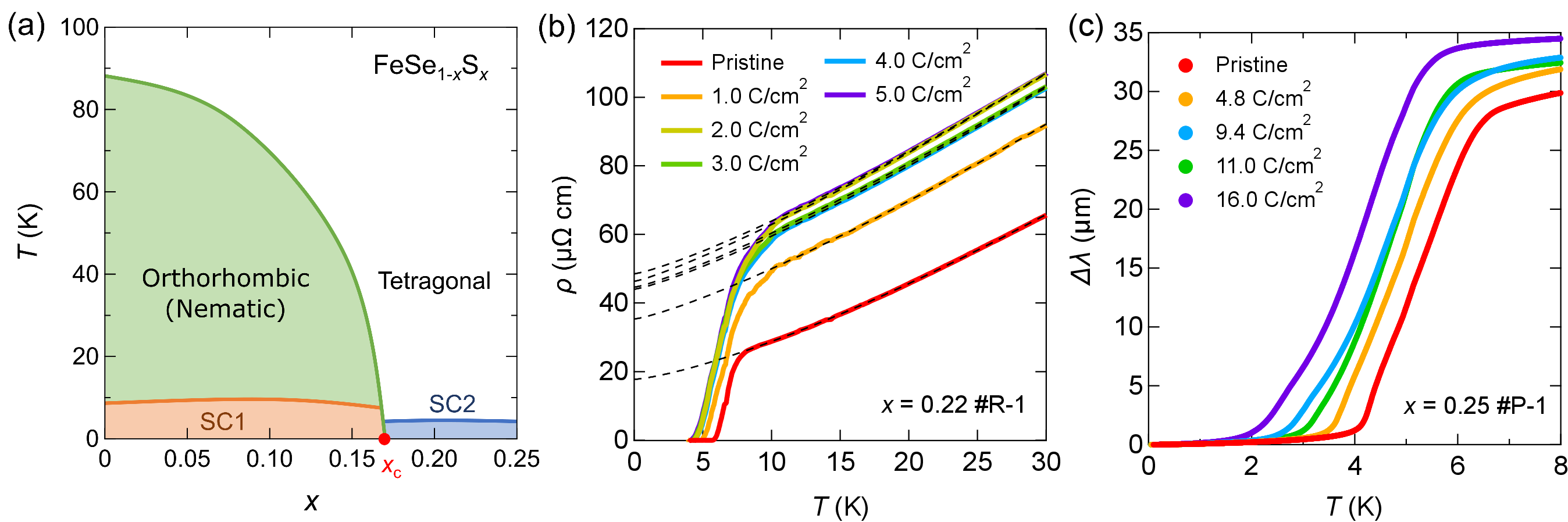}
    \caption{(a) Phase diagram of FeSe$_{1-x}$S$_x$ for $0\le x\le 0.25$. The nonmagnetic nematic phase (green region) disappears at the critical point $x_{\rm c}\approx 0.17$. The superconducting phases in the orthorhombic (orange) and tetragonal (blue) regions are denoted as SC1 and SC2, respectively. (b) Temperature dependence of the electrical resistivity below 30\,K for various irradiation doses in sample \#R-1 ($x=0.22$). Black dashed curves represent an extrapolation of resistivity data down to zero temperature obtained by a power-law fitting from 10 to 30\,K. (c) Temperature dependence of the change in the penetration depth $\Delta\lambda(T)$ below 8\,K for various irradiation doses in sample \#P-1 ($x=0.25$).}
    \label{Fig1}
\end{figure*}

FeSe is a unique iron-based superconductor in that the electronic nematic order appears without any magnetic orders at ambient pressure\,\cite{shibauchi2020,sun2016,matsuura2017}. This nematic order is suppressed by the substitution of the Se sites with isovalent S atoms and as shown in Fig.\,\ref{Fig1}(a), FeSe$_{1-x}$S$_x$ shows a nematic quantum critical point (QCP) at $x_{\rm c}\approx 0.17$\,\cite{matsuura2017,hosoi2016,licciardello2019,huang2020}. In FeSe$_{1-x}$S$_x$, several exotic superconducting properties have been reported, such as Bardeen-Cooper-Schrieffer (BCS) to Bose-Einstein condensation (BEC) crossover\,\cite{Kasahara2016,hashimoto2020,mizukami2023}, giant superconducting fluctuations\,\cite{Kasahara2016,mizukami2023}, orbital-selective pairing state\,\cite{sprau2017}, and field-induced Fulde-Ferrell-Larkin-Ovchinnikov state\,\cite{kasahara2014,Kasahara2020,Kasahara2021}. Furthermore, recent muon spin relaxation ($\mu$SR) measurements have revealed time-reversal symmetry breaking (TRSB) below the superconducting transition temperature ($T_{\rm c}$) in both orthorhombic (SC1) and tetragonal (SC2) phases\,\cite{matsuura2023}.

It has also been reported that the superconducting states SC1 and SC2 are quite different\,\cite{sato2018,hanaguri2018,matsuura2017,mizukami2023}, despite that the normal-state electronic structure changes smoothly across the nematic QCP\,\cite{hanaguri2018,coldea2019}. In the SC1 phase, quite anisotropic superconducting gap structure with deep gap minima or line nodes is expected\,\cite{kasahara2014,sprau2017,sato2018,hashimoto2018}, and a sign reversal of the gap function between hole and electron bands is reported\,\cite{sprau2017,moore2015}. In the SC2 phase, on the other hand, specific heat, scanning tunneling microscopy (STM), and $\mu$SR measurements show large residual density of states (DOS) even at low temperatures\,\cite{hanaguri2018,sato2018,mizukami2023,matsuura2023}.
As an explanation for the peculiar residual DOS in the tetragonal FeSe$_{1-x}$S$_x$, an ultranodal superconducting state with two-dimensional gap nodes, called Bogoliubov Fermi surfaces (BFSs), has been theoretically proposed\,\cite{setty2020NC,setty2020PRB}. Recent photoemission and nuclear magnetic resonance experiments support the presence of BFSs in the tetragonal FeSe$_{1-x}$S$_x$\,\cite{nagashima2022,yu2023}.

Previously, the gap structures have been classified into three types: fully gapped, point nodal, and line nodal. In nodal superconductors, the field-induced low-energy DOS has been observed due to the Doppler energy shift of quasiparticles known as the Volovik effect\,\cite{Volovik1993}. A new class of ultranodal state with BFSs can be viewed as the spontaneous Volovik effect on the gap nodes, in which the pseudomagnetic field due to TRSB inflates the gap nodes into BFSs, leading to the residual DOS even at zero external field. It has been theoretically shown that in centrosymmetric superconductors, the inflated BFSs are protected by a topological invariant characterized by the sign of the Pfaffian of the unitarily transformed Bogoliubov-de Gennes Hamiltonian\,\cite{agterberg2017,brydon2018}. However, the original gap nodes before the inflation may not be symmetry-protected as in the case of $s_\pm$ state in iron-based superconductors with accidental nodes that can be lifted by disorder\,\cite{Mishra2009,mizukami2014}. This raises an intriguing question on the robustness of the ultranodal state against impurity scattering in FeSe$_{1-x}$S$_x$.

Here, we investigate the impurity effects on the superconducting state in tetragonal FeSe$_{1-x}$S$_x$ from penetration depth measurements. Nonmagnetic impurities are introduced via electron irradiation, which is known as an effective method to create uniform point defects without changing electronic and crystal structures significantly\,\cite{mizukami2014,Roppongi2023}. We show that the penetration depth at low temperatures can be fitted with a model based on the BFSs before irradiation. With increasing the irradiation dose, the temperature dependence of penetration depth $\lambda(T)$ shows a nonmonotonic change, indicative of the transition from the nodal to a nodeless and then to a gapless behavior with the impurity-induced Andreev bound state. These results suggest that the BFSs may be lifted by disorder in tetragonal FeSe$_{1-x}$S$_x$.

%Methods
Single crystals of FeSe$_{1-x}$S$_x$ in the tetragonal phase ($0.18\le x \le 0.25$) are grown by the chemical vapor transport method, and S composition $x$ is determined by the $c$ axis lattice constant measured by X-ray diffraction. Electron irradiation with the incident energy of 2.5\,MeV was performed on the SIRIUS Pelletron accelerator in Ecole Polytechnique. The electrical resistivity is measured by the four-terminal method with current applied within the $ab$ plane. The penetration depth is measured by the tunnel diode oscillator technique at $\sim13.8$\,MHz. The change in the penetration depth as a function of temperature $\Delta\lambda(T)=\lambda(T)-\lambda(0)$ is determined from the frequency shift $\Delta f(T)=f(T)-f(0)$. In this technique, we apply a weak ac magnetic field of the order of $\mu$T, which is much lower than the lower critical field.

%Results
Figure\,\ref{Fig1}(b) shows the temperature dependence of the in-plane resistivity $\rho$ for $x=0.22$ before and after successive electron irradiation. As the irradiation dose increases, the residual resistivity $\rho_0$ is enhanced, indicating that the impurity scattering is successfully controlled. We note that the increase of $\rho_0$ is not proportional to the irradiation dose, which may reflect the annealing effect of the Frenckel pairs at high temperatures in this material. The normal-state $\rho(T)$ curves after irradiation show almost parallel shifts up to $\sim 150$\,K (see Fig. 1(b) and Fig. S1), implying that the electronic structure does not change against irradiation. Another feature is that $T_{\rm c}$ is clearly suppressed as the impurity scatterings are introduced. This suppression of $T_{\rm c}$ is in contrast to conventional superconductors, where $T_{\rm c}$ is insensitive to nonmagnetic impurities according to the Anderson theorem. This impurity effect on $\rho(T)$ is reproduced in another sample with $x=0.18$ (see Fig.\,S1).

Figure\,\ref{Fig1}(c) represents $\Delta\lambda(T)$ measured in another sample ($x=0.25$) with successive irradiation. We determine $T_{\rm c}$ as the onset temperature of the increase of normalized superfluid density $\rho_{\rm s}(T)=\lambda^2(0)/\lambda^2(T)$ (see Fig.\,S4). Near $T_{\rm c}$, a broad change of $\Delta\lambda(T)$ is observed in both the pristine and irradiated data. This gradual transition is unlikely to be induced by inhomogeneity of samples because we detected no signature of phase seperation or inhomogeneous conpositional distributions (see Fig.\,S3). Considering that similar broad transitions are also reported in the specific heat in the tetragonal FeSe$_{1-x}$S$_x$, the broad change of $\Delta\lambda(T)$ is likely associated with the strong superconducting fluctuations in the BCS-BEC crossover regime\,\cite{mizukami2023}.
The suppression of $T_{\rm c}$ by irradiation is also found in $\Delta\lambda(T)$. As $T_{\rm c}$ decreases, the apparent $\Delta\lambda(T)$ value in the normal state increases. This reflects an increase in the skin depth $\delta=\sqrt{\rho/\pi f\mu}$, where $\mu$ is the permeability of the sample, due to the enhancement of the resistivity. These results confirm the systematic increase of $\rho$ and decrease of $T_{\rm c}$ induced by electron irradiation.

\begin{figure}[t]
\includegraphics[width=0.8\linewidth]{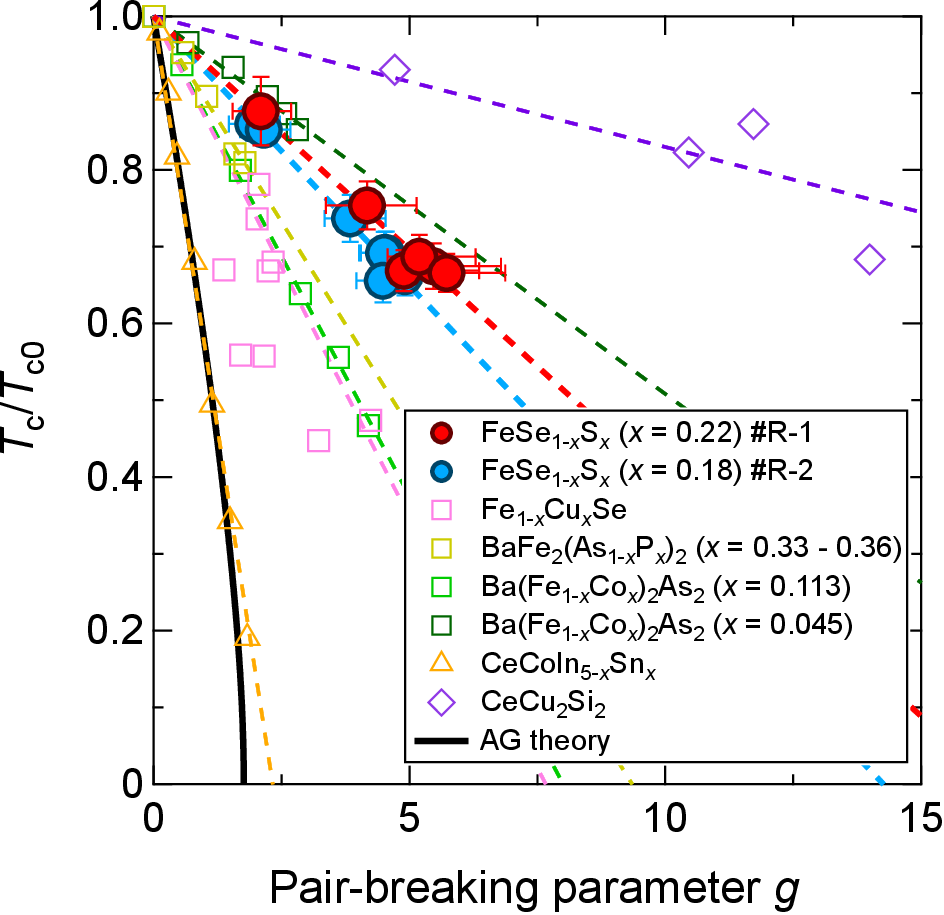}
\caption{Normalized superconducting transition temperature $T_{\rm c}/T_{\rm c0}$ in electron-irradiated FeSe$_{1-x}$S$_x$ as a function of pair breaking parameter $g=\hbar/\tau_{\rm imp}k_{\rm B}T_{\rm c0}$ for $x=0.22$ (red circles) and $x=0.18$ (blue circles). The black solid line represents a theoretical curve of the Abrikosov-Gor'kov (AG) theory. For comparison, the results of Sn-substituted CeCoIn$_5$ (orange triangles)\,\cite{bauer2006} are plotted as an example of $d$-wave superconductors. The results of other iron-based superconductors with $s_{\pm}$-wave symmetry such as Fe$_{1-x}$Cu$_x$Se (pink squares)\,\cite{Zajicek2022}, electron-irradiated BaFe$_2$(As$_{1-x}$P$_x$)$_2$ (yellow squares)\,\cite{mizukami2014}, proton-irradiated Ba(Fe$_{1-x}$Co$_{x}$)$_2$As$_2$ for $x=0.113$ (light green squares), and $x=0.045$ (dark green squeres)\,\cite{nakajima2010} are also shown. The results of electron-irradiated CeCu$_2$Si$_2$ (purple diamonds)\,\cite{yamashita2017} are plotted as an example of full-gap superconductors.}
\label{Fig2}
\end{figure}

To discuss the superconducting gap structure, we first focus on the quantitative analysis of the $T_{\rm c}$ suppression rate. As an index representing the relative strength of impurity scattering against the superconducting energy scale, we use the pair-breaking parameter $g=\hbar/\tau_{\rm imp}k_{\rm B}T_{\rm c0}$, where $\tau_{\rm imp}=m^*/ne^2\rho_0$ and $\rho_0$ is the residual resistivity. The value of $m^*/n$ is estimated from the quantum oscillation measurements\,\cite{coldea2019} (see Supplemental Materials), and $T_{\rm c0}$ is determined by an extrapolation of $T_{\rm c}$ to the $\rho_0\to0$ ($g \to 0$) limit. Figure\,\ref{Fig2} shows the suppression of $T_{\rm c}$ as a function of $g$ in tetragonal FeSe$_{1-x}$S$_x$ and several other superconductors. Although the $T_{\rm c}$ suppression in tetragonal FeSe$_{1-x}$S$_x$ is slower than the well-known Abrikosov-Gor'kov (AG) theory, $T_{\rm c}$ is more sensitive than that of a full-gap superconductor CeCu$_2$Si$_2$\,\cite{yamashita2017}. The fact that the $T_{\rm c}$ suppression rate in FeSe$_{1-x}$S$_x$ is as high as those in other iron-based superconductors with highly anisotropic superconducting gap implies that the superconducting gap in tetragonal FeSe$_{1-x}$S$_x$ is also highly anisotropic. Our results are in line with the recent theoretical calculations on the disorder effects, showing that the initial slopes of the $T_{\rm c}$ suppression for BFSs and line nodal cases are similar\,\cite{Oh2021}.

\begin{figure}[t]
\includegraphics[width=1.0\linewidth]{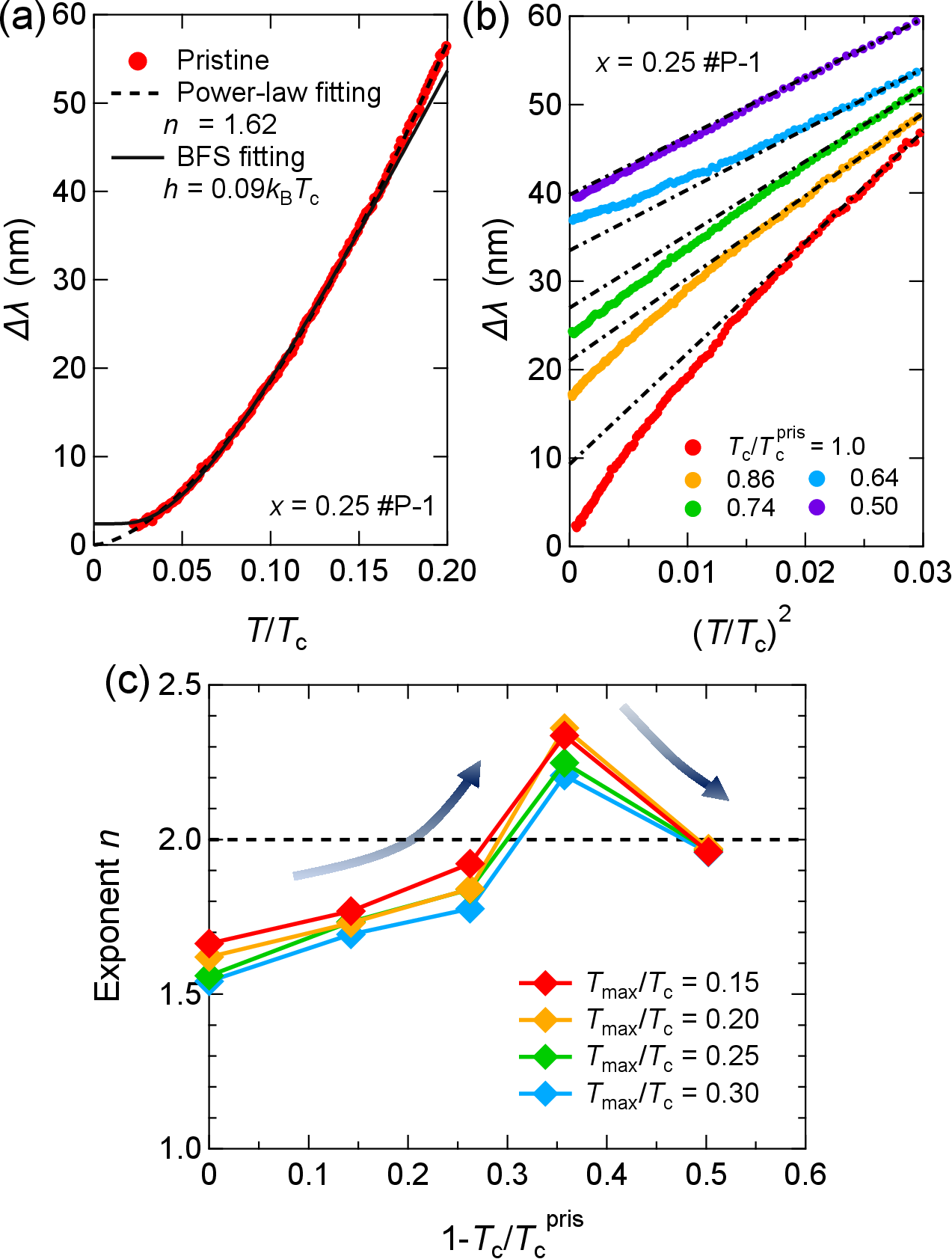}
\caption{(a) Temperature dependence of the penetration depth below $T/T_{\rm c}=0.2$ for the pristine sample with $x=0.25$. The black dashed line and solid line represent fitting curves by the power-law and the BFS model, respectively. The data points are vertically shifted so that the power-law fitting starts at $\Delta \lambda(0)=0$. (b) Change in the penetration depth as a function of $(T/T_{\rm c})^2$ for various irradiation doses. The data of irradiated samples are shifted vertically for clarity. Black dashed-dotted lines represent a linear function for the eye guide. (c) Exponent $n$ obtained from the power-law fitting as a function of $1-T_{\rm c}/T_{\rm c}^{\rm pris}$. $T_{\rm max}$ is the upper limit of the fitting temperature range.}
\label{Fig3}
\end{figure}

\begin{figure*}[t]
\includegraphics[width=0.8\linewidth]{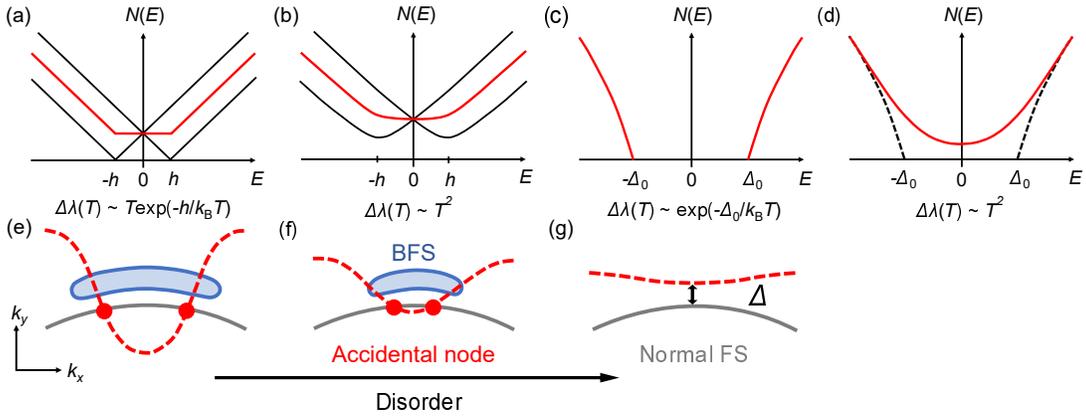}
\caption{(a)-(d) Schematic energy dependence of quasiparticle DOS, $N(E)$, for various nonmagnetic impurity concentrations. (a) $N(E)$ of the BFS model in the clean limit\,\cite{lapp2020}. (b) $N(E)$ of the BFS state with small impurity concentrations. (c) $N(E)$ in a fully gapped state with gap minima $\Delta_0$. (d) $N(E)$ in a gapless state with Andreev bound states. (e)-(g) Schematic pictures of a possible evaluation of the gap structure induced by disorder. Gray curves represent a part of normal-state Fermi surfaces in the $k_x$-$k_y$ plane. Red dashed lines show the momentum dependence of the original gap (without TRSB) relative to the Fermi surface. Red dots are the positions of accidental line nodes where the original gap size becomes zero. Blue solid curves depict BFSs in the $k_x$-$k_y$ plane induced by TRSB. BFSs surround the regions with negative Pfaffian (blue shades). The gap structures described by (e)-(g) correspond to $N(E)$ shown in (a)-(c), respectively.}
\label{Fig4}
\end{figure*}

Next, we examine the impurity effects on the low-temperature penetration depth in FeSe$_{1-x}$S$_x$, which provides pivotal information on the gap structure. To analyze the temperature dependence, we first apply a power-law fitting $\Delta\lambda\propto T^n$ below $0.2T_{\rm c}$, as depicted in Fig.\,\ref{Fig3}(a). The exponent $n$ plays a crucial role in discussing the gap structure because line nodes give $n=1$ in the clean limit, and with increasing impurity scatterings, $n$ increases up to the dirty limit value of $n=2$. To visualize the change in $\Delta\lambda(T)$ induced by impurities, we plot $\Delta\lambda(T)$ as a function of $(T/T_{\rm c})^2$ in Fig.\,\ref{Fig3}(b). Here, $T_{\rm c}/T_{\rm c}^{\rm pris}$ is used as an index of the strength of impurity scatterings instead of the irradiation doses, where $T_{\rm c}^{\rm pris}$ is $T_{\rm c}$ of the pristine sample. The pristine, $T_{\rm c}/T_{\rm c}^{\rm pris}=0.86$, and 0.74 samples show a downward deviation from the linear function at low temperatures, reflecting the exponent $n<2$. As $T_{\rm c}$ decreases, in contrast, the $T_{\rm c}/T_{\rm c}^{\rm pris}=0.64$ sample shows an upward deviation, meaning $n>2$ in this sample. Finally, the $T_{\rm c}/T_{\rm c}^{\rm pris}=0.50$ sample depicts a linear function, {\it i.e.} $n\approx 2$. Therefore, Fig.\,\ref{Fig3}(b) clearly indicates that $n$ changes nonmonotonically with nonmagnetic impurities. The exponent values obtained from the power-law fitting for $\Delta\lambda(T)$ of the pristine and irradiated samples are plotted against $1-T_{\rm c}/T_{\rm c}^{\rm pris}$ in Fig.\,\ref{Fig3}(c). This nonmonotonic change is independent of the choices of fitting range, as shown in Fig.\,\ref{Fig3}(c).

In superconductors with symmetry-protected nodes, like $d$-wave superconductors, the exponent $n$ changes from 1 to 2 monotonically\,\cite{Hirschfeld1993}. Therefore, our results showing a nonmonotonic change in $n$ rule out the possibilities of symmetry-protected nodes. A similar nonmonotonic change of $n$ induced by electron irradiation is also seen in BaFe$_2$(As$_{1-x}$P$_x$)$_2$, which has been explained by an $s_{\pm}$-wave gap with accidental nodes\,\cite{mizukami2014}. In general, accidental nodes can be lifted due to the gap-averaging effects of impurity scatterings, and then a fully gapped state may appear\,\cite{Mishra2009}. As more impurities are introduced, Andreev bound states are created by interband impurity scatterings for sign-changing order parameters, and the exponent $n$ becomes 2 in the dirty limit. Thus, the present results of $\Delta\lambda(T)$ are seemingly consistent with the $s_{\pm}$-wave gap with accidental nodes. As mentioned above, however, recent experiments reported large residual DOS and TRSB in the superconducting state, which cannot be accounted for by the simple $s_{\pm}$-wave state. Furthermore, our results of $\Delta\lambda\propto T^{1.6}$ in the pristine sample is incompatible with the residual DOS of $\sim30$\% of the Sommerfeld coefficient (Fig.\,S5) in a line-node model, in which $\Delta\lambda\propto T^2$ is expected. Therefore, we will discuss the possible impurity effects on the BFS state below.

When line nodes are inflated into BFSs, the energy-linear DOS, $N(E)$, in the clean limit is shifted by the spontaneous Volovik effect creating a flat part in $N(E)$ at low energies, as shown by the red line in Fig.\,\ref{Fig4}(a). In this simple model, the BFS state leads to $\Delta\lambda(T)\propto T\exp(-h/k_{\rm B}T)$, where $h$ is an effective pseudomagnetic field associated with TRSB\,\cite{lapp2020}. This temperature dependence fits well with the pristine data at low temperatures, as shown by the solid line in Fig.\,\ref{Fig3}(a). Now we extend the BFS model to the case with impurity scatterings by adding energy-shifted DOS as $N(E)\propto (E+h)^2/(|E+h|+E_{\rm imp})+(E-h)^2/(|E-h|+E_{\rm imp})$, as shown in Fig.\,\ref{Fig4}(b). This is based on a consideration that for the original line nodes, the impurity scattering changes from $N(E)\propto |E|$ to $N(E)\propto E^2/(|E|+E_{\rm imp})$\,\cite{Hirschfeld1993}, where $E_{\rm imp}$ is an effective strength of the impurity scatterings. Then, the inflation of line nodes into BFSs results in the red curve in Fig.\,\ref{Fig4}(b). However, the nonmonotonic behavior of $\Delta \lambda (T)$ cannot be reproduced by only this impurity effect as discussed in Supplemental Materials. Moreover, recent calculations for the BFS state\,\cite{Miki2021,Miki2024} suggest that nonmagnetic impurities can induce in-gap DOS reminiscent of bound states showing $n=2$, which, however, may not account for the overshoot ($n>2$).

Let us discuss the stability of the BFSs in tetragonal FeSe$_{1-x}$S$_x$. The obtained nonmonotonic change of $n$ can rule out that the nodal structure is protected by the symmetry of the order parameter. Thus, in the pristine sample, the original line nodes should also be accidental ones, which are then inflated into the BFS due to TRSB, as schematically shown in Fig.\,\ref{Fig4}(e). As the impurity scattering is introduced, the original gap structure becomes more isotropic (Fig.\,\ref{Fig4}(f)), and finally, the original accidental nodes can be lifted (Fig.\,\ref{Fig4}(g)). Considering that the inflated BFSs are located near the original nodes, BFSs can also disappear when the impurity scattering becomes strong enough. This scenario naturally explains that nonmagnetic impurities can change the gap structure from a BFS state to a fully gapped state. Following this evolution, $\Delta\lambda(T)$ changes from $T\exp(-h/k_{\rm B}T)$ to $T^n$ with $n\sim2$, and then to $\exp(-\Delta_0/k_{\rm B}T)$, which can be mimicked by $T^n$ with $n>2$, as shown in Fig.\,\ref{Fig4}(a)-(c). When the sign of gap function changes in different bands, the Andreev bound states are created even in fully gapped states, and $\Delta\lambda(T)\propto T^2$ is expected in the dirty limit (Fig.\,\ref{Fig4}(d)). Our results of $\Delta\lambda(T)\propto T^2$ in the highly-irradiated sample indicate that the gap function has a sign reversal. Therefore, we can qualitatively explain the observed nonmonotonic dependence of $\Delta\lambda(T)$ from the BFS scenario described above. We emphasize here that the observed sign-changing nature of gap function is also crucial in the BFS model because the criteria of whether the BFSs appear or not depends on the sign of gap function on each band\,\cite{setty2020NC,setty2020PRB}. For more quantitative analyses of our data, microscopic models of impurity effects on the BFS state are highly desired based on the electronic structures of tetragonal FeSe$_{1-x}$S$_x$.

Finally, we comment on another theory recently proposed to explain the unusual large residual DOS in FeSe$_{1-x}$S$_x$ based on the nematic quantum critical fluctuations\,\cite{Islam2024}. In this theory, a rapid downturn behavior of $\Delta\lambda(T)$ is expected below a characteristic temperature which strongly depends on the closeness to the nematic QCP. However, $\Delta\lambda(T)$ data for two different compositions $x=0.18$ and 0.25 both show no downturn down to $\sim 0.02T_{\rm c}$ (see Fig.\,S7). Therefore, we consider that the ultranodal state with accidental BFSs is the most plausible scenario to consistently explain the unusual superconducting properties in the tetragonal FeSe$_{1-x}$S$_x$.

In summary, we investigated impurity effects on superconductivity in the ultranodal superconductor candidate FeSe$_{1-x}$S$_x$ using electron irradiation. With increasing impurity scattering, $T_{\rm c}$ is suppressed, and the low-energy quasiparticle excitations detected by penetration depth measurements show a nonmonotonic behavior, evidencing transitions of the gap structure. Our results highlight unusual impurity effects on BFSs, which point to accidental nature of the nodes in FeSe$_{1-x}$S$_x$ leading to the disorder-induced lifting of BFSs.

We thank P.\ J.~Hirschfeld, S.~Kitou, and E.-G.~Moon for fruitful discussions.
This work was supported by Grants-in-Aid for Scientific Research (KAKENHI) (Nos.\ JP22H00105, JP22K20349, JP21H01793, JP19H00649, JP18H05227, JP18KK0375), Grant-in-Aid for Scientific Research on innovative areas ``Quantum Liquid Crystals" (No.\ JP19H05824), Grant-in-Aid for Scientific Research for Transformative Research Areas (A) ``Condensed Conjugation" (No.\ JP20H05869) from Japan Society for the Promotion of Science (JSPS), CREST (No.\ JPMJCR19T5) from Japan Science and Technology (JST), and LabEx PALM (ANR-10LABX-0039-PALM). The authors acknowledge support from the EMIR\&A French network (FR CNRS 3618) on the platform SIRIUS.

%\bibliography{refs_ver4}  

\end{document}